\documentclass[aps,prl,reprint,superscriptaddress,doi=false,isbn=false,url=false,title=true,longbibliography,noeprint]{revtex4-2}
\usepackage{hyperref}
\hypersetup{colorlinks=false, citecolor=blue, urlcolor=blue, linkcolor=blue}
\usepackage{amsmath}
\usepackage{amssymb}
\usepackage{float}
\usepackage{graphicx}
\usepackage{xcolor}

\begin{document}
\title{Non-universality of aging during phase separation of the two-dimensional long-range Ising model}

\affiliation{Institut f\"ur Theoretische Physik, Universit\"at Leipzig, IPF 231101, 04081 Leipzig, Germany}
\affiliation{NEC Laboratories Europe GmbH, Kurfürsten-Anlage 36, 69115 Heidelberg, Germany}
\author{Fabio M\"uller}
\email{fabio.mueller@itp.uni-leipzig.de}
\affiliation{Institut f\"ur Theoretische Physik, Universit\"at Leipzig, IPF 231101, 04081 Leipzig, Germany}
\author{Henrik Christiansen}
\email{henrik.christiansen@neclab.eu}
\affiliation{Institut f\"ur Theoretische Physik, Universit\"at Leipzig, IPF 231101, 04081 Leipzig, Germany}
\affiliation{NEC Laboratories Europe GmbH, Kurfürsten-Anlage 36, 69115 Heidelberg, Germany}
\author{Wolfhard Janke}
\email{wolfhard.janke@itp.uni-leipzig.de}
\affiliation{Institut f\"ur Theoretische Physik, Universit\"at Leipzig, IPF 231101, 04081 Leipzig, Germany}
\date{\today}

\graphicspath{{figures/}}

\begin{abstract}
  We investigate the aging properties of phase-separation kinetics following quenches from $T=\infty$ to a finite temperature below $T_c$ of the paradigmatic two-dimensional conserved Ising model with power-law decaying long-range interactions $\sim r^{-(2 + \sigma)}$.
  Physical aging with a power-law decay of the two-time autocorrelation function $C(t,t_w)\sim \left(t/t_w\right)^{-\lambda/z}$ is observed, displaying a complex dependence of the autocorrelation exponent $\lambda$ on $\sigma$.
  A value of $\lambda=3.500(26)$ for the corresponding nearest-neighbor model (which is recovered as the $\sigma \rightarrow \infty$ limes) is determined.
  The values of $\lambda$ in the long-range regime ($\sigma < 1$) are all compatible with $\lambda \approx 4$.
  In between, a continuous crossover is visible for $1 \lesssim \sigma \lesssim 2$ with non-universal, $\sigma$-dependent values of $\lambda$.
  The performed Metropolis Monte Carlo simulations are primarily enabled by our novel algorithm for long-range interacting systems.
\end{abstract}

\maketitle
Phase separation constitutes one of the most fundamental processes underlying the ordering process during the relaxation of systems towards equilibrium.
It can be observed in completely diverse physical settings from quantum systems~\cite{Hoffer1986,ao2000two,chomaz2004nuclear,hofmann2014coarsening} and biophysics~\cite{fan2010hydrodynamic,marchetti2013hydrodynamics,alberti2017phase,jiang2020protein,yoshizawa2020biological} to cosmology~\cite{boyanovsky2006phase,binney2011galactic}.
Beyond the mere theoretical interest~\cite{bray2002theory,puri2009kinetics,Henkel2010,livi2017nonequilibrium} the involved mechanisms are highly relevant to industrial applications~\cite{chomaz2004nuclear,hyman2014liquid,shimizu2015novel}.
From a computational point of view, the process of phase-separation kinetics attracted widespread attention and there are numerous works by several groups which investigate different aspects of this complex process for systems with short-range interactions~\cite{Weinkamer2004,Majumder2010,Ahmad2010,Singh2014,Lu2020,Joseph2021,Bhattacharyya2024}.
\par
Many realistic physical circumstances, however, involve long-range potentials~\cite{eyink2006onsager,french2010long,levin2014nonequilibrium,campa2014physics,douglas2015quantum,neyenhuis2017observation,zhang2018long,hazra2021biophysics}.
Such systems have received much less attention in computer simulation studies since their treatment is prohibitively expensive without efficient algorithms.
Recently, we have developed a new algorithm for fast Monte Carlo simulation of long-range interacting systems~\cite{Mueller2023}, which puts us now in the position to study the phase-separation kinetics also in presence of long-range interactions for sufficiently large systems and observation times.
\par
Here, we focus on the dynamics of phase separation using the paradigmatic long-range Ising model (LRIM) with conserved order parameter (COP) dynamics, whose Hamiltonian is given by
\begin{equation}
  \mathcal{H}= - \frac{1}{2}\sum_i\sum_{j \neq i}J_{i,j}s_is_j,\quad J_{i,j}=r_{i,j}^{-(d+\sigma)},
  \label{eq:hamiltonian}
\end{equation}
where the spins take values $s_i=\pm 1$, $r_{i,j}$ is the distance between $s_i$ and $s_j$, $d$ is the dimension, and the parameter $\sigma$ describes the spatial decay of the interactions being reflected in the spin-spin couplings $J_{i,j} > 0$.
We consider two-dimensional $L\times L$ square lattices for which the periodic boundary conditions are implemented by a pre-calculated Ewald summation~\cite{Ewald1921} in combination with the minimum-image convention.
For different values of $\sigma$ the system's behavior will fall into distinct nonequilibrium universality classes concerning the temporal growth of the characteristic length scale $\ell(t)\sim t^\alpha$ with $\alpha=1/z$, which can be thought of as the average linear size of the magnetic domains at time $t$.
In Ref.~\cite{Mueller2022} we observed for the $d=2$ case the predicted continuous crossover~\cite{bray1993domain,Bray1994,rutenberg1995energy} from a long-range regime ($\sigma < 1$) with $\sigma$-dependent growth exponent $\alpha = 1/(2+\sigma)$ to a short-range regime ($\sigma>1$) with a $\sigma$-independent exponent $\alpha=1/3$ equal to the nearest-neighbor (NN) case (corresponding to $\sigma=\infty$).
This is similar to the behavior of the LRIM with non-conserved order parameter (NCOP) dynamics where the corresponding predictions~\cite{bray1993domain,Bray1994,rutenberg1995energy} have been confirmed numerically~\cite{corberi2019one,Christiansen2019a}: The growth of $\ell(t)$ is described by the NN-exponent $\alpha=1/2$ for $\sigma > 1$ and by a $\sigma$-dependent exponent $\alpha = 1/(1+\sigma)$ for $\sigma<1$, being again continuous at $\sigma=1$.
\par
\begin{figure}
  \includegraphics{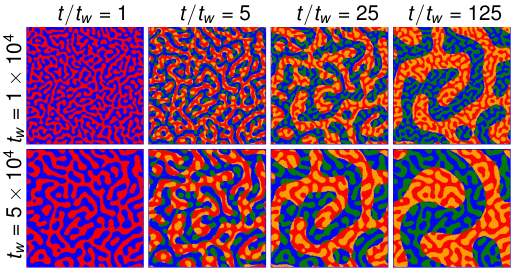}\\
  \caption{Evolution of the local autocorrelation for $\sigma = 0.8$ and $L=1024$, each row for a different references time $t_w$ and each column for a constant value of the scaling variable $y=t/t_w$.
    The red (blue) spins point up (down) at times $t_w$  \emph{and} at $t$ and contribute with a positive sign to the autocorrelation function.
    The green (orange) spins point up (down) at times $t_w$  \emph{but} down (up) at $t$, thus,  entering with a negative sign.
    The growth of the average domain sizes (size of the patches of red + orange and blue + green, respectively) as well as the decay of the autocorrelation (shrinking area of blue + red) with increasing time from left to right are clearly visible.
  }
  \label{fig:aging_config}
\end{figure}
Another fundamental aspect of phase-ordering processes is the behavior of two-time quantities, such as the autocorrelation function 
\begin{equation}
  \label{eq:two-time-correlation}
  C(t,t_w) = \langle s_i(t)s_i(t_w) \rangle - \langle s_i(t) \rangle  \langle s_i(t_w) \rangle,
\end{equation}
which measures the similarity of the system at time $t$ to some reference time $t_w<t$.
The operator $\langle\ldots\rangle$ is intended as an average over different quench experiments, i.e.,  pairs of initial configurations and realizations of thermal noise.
The second term in this definition does not contribute since we are performing conserved order-parameter simulations with magnetization $M = \sum_i s_i = 0$ (often referred to as critical mixture) and, hence, trivially $\langle s_i \rangle = 0$.
In particular, as common for phase ordering, we expect physical aging~\cite{Henkel2010} where dynamical scaling of the autocorrelation function is observed (becoming a function of $y=t/t_w$) and its asymptotic decay 
\begin{equation}
  \label{eq:two-time-powerlaw}
  C(t,t_w) \sim (t/t_w)^{-\lambda/z}
\end{equation} 
is described by a power law with an additional independent exponent, the autocorrelation exponent $\lambda$.
\par
For instance, aging is observed for the two-dimensional LRIM with NCOP in Ref.~\cite{Christiansen2020}.
There, for all considered $\sigma$ the values of $\lambda$ are compatible with a jump of $\lambda$ at $\sigma=1$ from a value $\lambda \approx 1 = d/2$ for $\sigma<1$ to a value of $\lambda \approx 1.25$ (equal to the NN case) for $\sigma > 1$.
That behavior is in good correspondence to the observations in the one-dimensional LRIM with NCOP~\cite{Corberi2019}, where $\lambda=0.5=d/2$ for $\sigma \leq 1$ and $\lambda = 1$ (again equal to the NN case) for $\sigma>1$ is found.
Hence, in both dimensions the autocorrelation exponent does not lead to any further splitting of the dynamical universality class, while marginally fulfilling the Fisher-Huse bound of $\lambda=d/2$~\cite{Fisher1988} for $\sigma<1$ and taking the value of the corresponding NN model for $\sigma > 1$.
\par
Aging is also observed in several phase-separating systems.
In Ref.~\cite{Midya2015} $\lambda \approx 3.6$ was observed for the NN Ising model with COP.
The long-range regime ($\sigma<1$) of the LRIM with COP was recently investigated in Ref.~\cite{Ghosh2024} where a non-trivial $\sigma$-dependence of $\lambda= c/\alpha=c (2+\sigma) $ with $c \approx 1.1$ was reported.
Such a slower decay of the autocorrelation function in the long-range regime would be in accordance to the observation for the NCOP case while the non-trivial $\sigma$-dependence would be a novel feature.
\par
By exploiting our recent algorithm~\cite{Mueller2023}, we aim in this Letter at a comprehensive quantitative assessment of the $\sigma$-dependence of $\lambda$ during phase separation in the LRIM with COP by performing Monte Carlo simulations for a large set of $\sigma$-values, ranging from the true NN-model ($\sigma = \infty$) to the long-range regime ($\sigma = 0.6$).
Our large-scale simulations on square lattices of size up to $2048^2$ and observation times up to about $10^8$ are compatible with a constant $\lambda \approx 4$ in the long-range regime $\sigma < 1$ and provide evidence for an extended crossover regime to the NN-limit $\lambda \approx 3.50$ from about $\sigma = 1$ to $\sigma=2$, coinciding with the range where also the \emph{equilibrium} critical exponents show a non-trivial $\sigma$-dependence \cite{Fisher1972,Sak1973}.
This suggests a possible, unexpected connection between aging properties and \emph{equilibrium} critical behavior.
\par
\begin{figure}
  \includegraphics{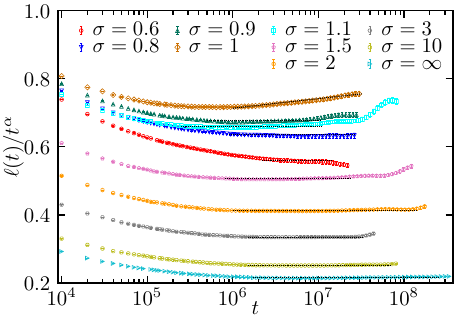}\\
  \caption{    
    Characteristic length $\ell(t)/t^\alpha$ together with the predicted growth law (dotted black lines) for all considered values of $\sigma$ (the system sizes differ among the different $\sigma$, see text).
    Error bars indicate twice the standard error of the mean.
    For $\sigma=0.6$ the observed growth matches the prediction for little less than a decade while for all other $\sigma$ the correspondence extends over $\approx 1.5-2$ decades.
  }
  \label{fig:ell}
\end{figure}
\par
For the simulation protocol and the implementation of the boundary conditions we follow our Ref.~\cite{Mueller2022}.
The phase-separation kinetics is studied by initially placing equally many up and down spins randomly on a square lattice ($M \equiv 0$).
Subsequently, the system is quenched to temperature $T_q=0.5 T_c$, well below the critical temperature $T_c$~\cite{Horita2017upper},  where it relaxes through local Monte Carlo Kawasaki dynamics~\cite{kawasaki1966diffusion}, i.e., spins are only allowed to exchange with their direct neighbors, which is the simplest local update scheme leaving the total magnetization conserved.
Getting sufficiently accurate estimates for $\lambda$ required us to simulate even larger system sizes than considered in Ref.~\cite{Mueller2022} in order to access later finite-size (FS) unaffected times and thus observe a longer asymptotic period.
In total, we performed 100 runs with different initial conditions and thermal noises for each $\sigma \leq 1$ with $L=2048$, 400 runs for $1.1 \leq \sigma \leq 2$  with $L=1024$, 1000 runs for $3 \leq \sigma \leq 10$ with $L=512$, and 400 runs for the NN case with $L=1024$ to push the FS time to more than $10^7$ for all $\sigma$, also extending and consolidating our findings for $\ell(t)$ in Ref.~\cite{Mueller2022}.
Despite the efficiency of our novel algorithm~\cite{Mueller2023}, the longest simulations ran up to $9$ months individually on a single CPU core, amounting to a total runtime of roughly $1000$ core years.
\par
\begin{figure}
   \includegraphics{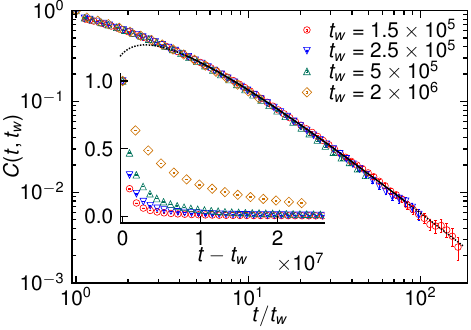}\\
  \caption{
    The autocorrelation function $C(t,t_w)$ for $\sigma=0.8$ and $L=2048$ for different waiting times $t_w$ shows an excellent data collapse when plotted against $t/t_w$ (main plot) while time-translation invariance is clearly violated (inset).
    The solid black line shows the best fit within the final chosen fitting range and the dotted black line its continuation to larger and smaller values of $y=t/t_w$ (see text for discussion).
  }
  \label{fig:tti-plot}
\end{figure}
After the quench the system relaxes towards equilibrium (being a phase-separated stripe state with some thermally excited spins in both phases) through the formation of domains which grow in time and whose average linear size is described by the characteristic length scale $\ell(t)$.
This phenomenon can be visually appreciated from Fig.~\ref{fig:aging_config} where one exemplary temporal evolution of the local autocorrelations for $\sigma=0.8$ and $L=1024$ is displayed.
The domains of up spins at time $t$ (red + orange) and the domains of down spins (blue + green) do clearly become larger with growing $t$.
A more quantitative analysis of the growth is obtained from $\ell(t)$ which we measure from the zero-intersect of the correlation function.
The representation $\ell(t)/t^{\alpha}$ chosen in Fig.~\ref{fig:ell} highlights the deviation from the expected growth law.
The extended periods of asymptotic growth (represented by the black dotted lines which for $\sigma=1$ incorporate the expected logarithmic correction~\cite{bray1993domain,Bray1994}) will be important for the discussion of the aging properties which we focus on in the following.
\par
In addition to the growth of $\ell(t)$, Fig.~\ref{fig:aging_config} also visualizes the evolution of the autocorrelation.
The time $t$ of an individual panel in the bottom row and the panel top right to it are always identical (and thus also the configuration).
Moving from left to right the amount of blue (red) decreases showing that the value of the autocorrelation function decreases.
Or alternatively, the position of the domains at time $t$ does correlate less and less with the domains at $t_w$ (leftmost panels).
Comparing the top row to the bottom row (different $t_w$ but same $y=t/t_w$) due to dynamical scaling and the conserved order parameter dynamics the total area of all four colors, individually, should be roughly the same.
The larger number of patches for the earlier $t_w$ is compensated by a larger patch size for the later one, altogether fulfilling the dynamical scaling hypothesis.
For physical aging to occur the relaxation dynamics needs to be slow (sub-exponential), $C(t,t_w)$ needs to become a function of the scaling variable $y=t/t_w$ (dynamical scaling) instead of being a function of $t-t_w$ (loss of time-translation invariance)~\cite{Henkel2010}. The latter two points are checked exemplary for $\sigma=0.8$ in Fig.~\ref{fig:tti-plot} where for different values of $t_w$ one observes a very good collapse of $C(t,t_w)$ when plotted as a function of $y$ in the main plot but no collapse is visible in the inset, where $t-t_w$ is chosen as putative scaling variable.
\par
Asymptotically for $t_w \gg 1$ and $y \rightarrow \infty$ the autocorrelation function is expected to follow a power law $C(t,t_w)\sim y^{-\lambda/z}$.
Since asymptotically $\ell(t)\sim t^\alpha=t^{1/z}$ this can be also understood as $C(t,t_w)\sim \left[ \ell(t)/\ell(t_w) \right]^{-\lambda} \equiv x^{-\lambda}$ and one may either take $x$ or $y$ as a scaling variable for the autocorrelation function.
If $t_w$ is not chosen in the asymptotic regime, however, the quality of the dynamical scaling is different for the two different scaling variables since the pre-asymptotic effects in $\ell(t)$ (cf. Fig.~\ref{fig:ell}) can accumulate and manifest in $x$.
In our setting, we find more consistent dynamical scaling in $y$ and, therefore, stick to it as scaling variable, since the asymptotic behavior of $\lambda$ should be unaffected by the choice of the scaling variable (see Supplemental Material~\footnote{See Supplemental Material at \{will be added\} where the raw data of $C(t,t_w)$ for all considered values of $\sigma$ are presented.
  We compare the quality of the dynamical scaling in plots using $y=t/t_w$ respectively $x=\ell(t)/\ell(t_w)$ and, extending over several figures and tables, show how the final values of $\lambda$ are determined from fits in $y$ (dependence of the fit parameters on the lower fit boundary $y_\mathrm{min}$, the finally used fit intervals, and the resulting final fit parameters).}).
\par
The autocorrelation exponent can either be extracted from the instantaneous exponent $\lambda_i\equiv d \ln C(t,t_w)/d \ln t$ or alternatively by performing a fit to $C(t,t_w)$.
In the former case the expectation is that $\lambda_i$ for large times reaches a plateau corresponding to its asymptotic value.
Since this is often not the case, it is common to plot $\lambda_i$ as a function of the inverse of the scaling variable and performing a (visual) extrapolation of (some subset of) the data to the ordinate~\cite{Fisher1988,Midya2015}.
This procedure implicitly assumes some functional form of the corrections in $\lambda_i$ (depending on the employed scaling variable and the scaling of the axes, see also the discussion in Ref.~\cite{gessert2024aging}) and relies crucially on choosing a valid subset of the data, which can be a non-trivial task, particularly due to the often rather noisy nature of the involved numerical derivative.
We hence opted for the latter alternative by directly fitting $C(t,t_w)$ with two different functional forms: \emph{i}) $f= A y^{-\lambda/z}$ and \emph{ii}) $f= A y^{-\lambda/z}\left( 1 - B/y \right)$.
Here, alternative \emph{i}) is in principle conceptually the cleanest solution since it works without any assumption of the functional form of the involved corrections.
Its main drawback is that the purely asymptotic regime in $C(t,t_w)$ (corresponding to a plateau in the $\lambda_i$) is usually too small (often even invisible) in order to allow an unbiased and accurate quantitative assessment of $\lambda$.
Our final results are thus extracted with fits to \emph{ii}) including a generic correction term which has been shown to be justified for short-range interacting models~\cite{Picone2004,Henkel2010} as well as for long-range interacting models~\cite{Christiansen2020}, albeit for NCOP dynamics (for details see Supplemental Material of Ref.~\cite{Christiansen2020}).
\par
During the fitting procedure the main challenge lies in finding the right fitting window $[y_\mathrm{min},y_\mathrm{max}]$ which excludes the pre-asymptotic period (which is not fully accounted for by the assumed correction) as well as the FS-affected regime.
After discarding the data which shows FS effects in $\ell(t)$, finding the right cutoff $y_\mathrm{max}$ is achieved by observing that the FS effects for different $t_w$ set in at the same time $t_\mathrm{max}$ which can, thus, be determined quite accurately by directly comparing $C(t,t_w)$ for a pair of two different values of $t_w$.
The value for $y_\mathrm{min}$ is determined by keeping the previously established $y_\mathrm{max}$ fixed and examining the response of the value of reduced chi-square $\chi^2/\mathrm{dof}$ of the fit to a variation of $y_\mathrm{min}$ where $\mathrm{dof}$ stands for the degrees of freedom.
We choose $y_\mathrm{min}$ such that it becomes minimal under the condition that its variation leaves the resulting $\chi^2/\mathrm{dof}$ small.
For all the considered cases (for the fits to form \emph{ii}), especially) the choice of $y_\mathrm{min}$ did not have a relevant influence on the resulting value of $\lambda$, compare the Supplemental Material where we present raw data for all values of $\sigma$ together with the fits, and tables containing the final fit parameters and intervals used for the fits and the dependence of the fit parameters on $y_\mathrm{min}$.
\begin{figure}
  \includegraphics[width=\columnwidth]{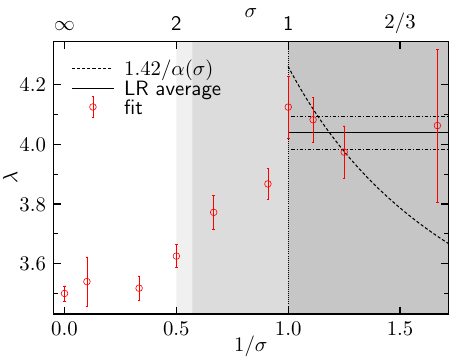}\\
  \caption{
    The autocorrelation exponent $\lambda$ plotted against the inverse of the exponent $\sigma$ of the power-law interaction.
    For large values of $\sigma$ we find $\lambda \approx 3.50$.
    In the long-range regime ($\sigma < 1$) the solid line represents the weighted average $4.038(56)$ with error bars indicated by the dash-dotted lines.
    The crossover happens continuously over an extended range $1 \lesssim\sigma\lesssim 2$.
    The lines refer to the nonequilibrium behavior while the shading of the background encodes the \emph{equilibrium} critical behavior (see text for discussion).
  }
  \label{fig:lambda-sigma}
\end{figure}
The best fit for $\sigma=0.8$ is shown in Fig.~\ref{fig:tti-plot} as a solid black line (the dotted line being the continuation of the fit outside the considered fit interval) which described the data over a very long interval very well.
\par
The full collection of obtained values of $\lambda$ from our analysis is plotted against  $1/\sigma$ in Fig.~\ref{fig:lambda-sigma}.
In the long-range regime with $\sigma<1$ we find that $\lambda \approx \mathrm{const} \approx 4$ (black solid line) is more conceivable than the hypothesis raised in Ref.~\cite{Ghosh2024} that $\lambda = c/\alpha = c(2+\sigma)$ (dashed line, where we manually adjusted the prefactor to $c=1.42$ instead of $c=1.1$ from Ref.~\cite{Ghosh2024} to be more plausible and fully fit into the plot).
For the NNIM, we find $\lambda = 3.500(26)$ which is similar albeit slightly smaller than the value previously reported in Ref.~\cite{Midya2015}.
In the white region ($\sigma$-independent NN \emph{equilibrium} critical exponents) $\lambda$ stays compatible with its NN value.
Over the (light) gray shaded area (where $\sigma$-dependent \emph{equilibrium} critical behavior is observed and/or expected~\cite{Fisher1972,Sak1973,Luijten2002,Picco2012,Blanchard2013,Angelini2014,Horita2017upper}) there is a continuous crossover to the long-range value $\lambda \approx 4$ visible in the dark gray region ($\sigma$-independent \emph{equilibrium} mean-field critical exponents).
At an intermediate value of $\sigma=1.5$ ($1/\sigma = 2/3$) the value of $\lambda \approx 3.745(48)$ is neither compatible with the NN value of $\lambda \approx 3.50$ nor with the long-range value of $\lambda\approx 4$.
\par
These findings are in contrast to what was observed for aging in the NCOP case~\cite{Christiansen2020} in two major ways: \emph{i}) There, $\lambda$ takes two distinct values, i.e., $\lambda=1$ in the long-range and $\lambda=1.25$ in the short-range regime with a more jump-like crossover behavior~\footnote{There, the spacing in $\sigma$ was wider than in this work but for $\sigma=0.6$ and $\sigma=0.8$ a value compatible to the long-range value and for $\sigma=1.5$ a value compatible with the short-range value of $\lambda$ was found}.
The location of the crossover, happening around $\sigma = 1$, coincides with the crossover from short- to long-range growth of $\ell(t)$~\cite{Bray1994,Christiansen2019a}.
Here, in the COP case, we find an extended crossover covering the intermediate range $1 \lesssim \sigma\lesssim 2$.
In this regime, where also the \emph{equilibrium} critical exponents depend non-trivially on $\sigma$, $\lambda$ assumes $\sigma$-dependent values, insinuating a possible influence of the former on the latter. 
\emph{ii}) Unexpectedly, the value of $\lambda\approx 4$ observed in the long-range regime is larger than $\lambda \approx 3.50$ observed for the NNIM case.
The phase-separation process is, thus, in a sense ``less efficient'' in presence of long-range interactions, since a faster decay of the autocorrelation function indicates that more ``mass'' has to be transported in order to achieve the same growth in $\ell(t)$.
This is in contrast to the reasoning in Ref.~\cite{Corberi2019} where it was argued that the asymptotically dominant, non-diffusive component of the motion of the domain walls, leading to an enhanced growth for $\sigma<1$, may also be responsible for the lower value of the autocorrelation exponent, thus, implying a ``more efficient'' ordering process in the long-range regime.
\par
To conclude, we have investigated aging and performed a quantitative analysis of the behavior of the autocorrelation function during the phase separation in the two-dimensional long-range Ising model.
The results were obtained by means of large-scale Metropolis Monte Carlo simulations enabled by our recent algorithm for the simulation of long-range interacting systems~\cite{Mueller2023}.
As expected, we find evidence of physical aging for all considered values of $\sigma$.
For the nearest-neighbor Ising model (corresponding to $\sigma = \infty$) we have determined the autocorrelation exponent $\lambda = 3.500(26)$ with unprecedented accuracy.
As in the case of phase ordering without conservation of the order parameter~\cite{Corberi2019,Christiansen2020} we observe a short- as well as a long-range regime with two distinct values of $\lambda$.
Intriguingly, we find compelling numerical evidence for a nontrivial and novel crossover behavior of $\lambda$ in the regime where also the \emph{equilibrium} critical exponents show a non-trivial $\sigma$-dependence.
\par
An interesting extension of this study could be quenches of off-critical mixtures or mixtures of a larger number of different components~\footnote{F. {M\"uller}, H. Christiansen, and W. Janke, in preparation (2024)}, since for sufficiently low concentration of the minority species only the transport of particles through evaporation and deposition on other droplets remains, for which the arguments presented in~\cite{corberi2019one} predicting a $\sigma$-independent growth law may be valid and, hence, also a different behavior of the autocorrelation function can be expected.

\begin{acknowledgments}
  This project was funded by the Deutsche Forschungsgemeinschaft (DFG, German Research Foundation) under project No. 189\,853\,844 -- SFB/TRR 102 (project B04), and the Deutsch-Französische Hochschule (DFH-UFA) through the Doctoral College ``$\mathbb{L}^4$'' under Grant No.\ CDFA-02-07. 
\end{acknowledgments}

\end{document}


\renewcommand{\thesection}{S\arabic{section}}
\renewcommand{\thetable}{S\arabic{table}}
\renewcommand{\thefigure}{S\arabic{figure}}

\def\theequation{S.\arabic{equation}}
\renewcommand{\figurename}{FIG.}

\title{Supplemental Material: Non-universality of aging during phase separation of the two-dimensional long-range Ising model}

\affiliation{Institut f\"ur Theoretische Physik, Universit\"at Leipzig, IPF 231101, 04081 Leipzig, Germany}
\affiliation{NEC Laboratories Europe GmbH, Kurfürsten-Anlage 36, 69115 Heidelberg, Germany}
\author{Fabio M\"uller}
\email{fabio.mueller@itp.uni-leipzig.de}
\affiliation{Institut f\"ur Theoretische Physik, Universit\"at Leipzig, IPF 231101, 04081 Leipzig, Germany}
\author{Henrik Christiansen}
\email{henrik.christiansen@neclab.eu}
\affiliation{Institut f\"ur Theoretische Physik, Universit\"at Leipzig, IPF 231101, 04081 Leipzig, Germany}
\affiliation{NEC Laboratories Europe GmbH, Kurfürsten-Anlage 36, 69115 Heidelberg, Germany}
\author{Wolfhard Janke}
\email{wolfhard.janke@itp.uni-leipzig.de}
\affiliation{Institut f\"ur Theoretische Physik, Universit\"at Leipzig, IPF 231101, 04081 Leipzig, Germany}
\date{\today}

\maketitle

This Supplemental Material contains the raw data for the autocorrelation function.
For these data the dynamical scaling for the two most common scaling variables $y=t/t_w$ and $x=\ell(t)/\ell(t_w)$ is presented.
Additionally, details on the fitting procedure for the chosen scaling variable $y$ are discussed, where in particular the choice of parameters of the fits and the resulting fitting parameters are presented.

Finally, the influence of the inclusion of an asymptotic correction term into the fit ansatz is investigated.

\section*{Loss of time-translation invariance and dynamical scaling}
\begin{figure*}
\subcaptionbox{$\sigma=0.6$ \label{0.6}}{\includegraphics[width=0.64\columnwidth]{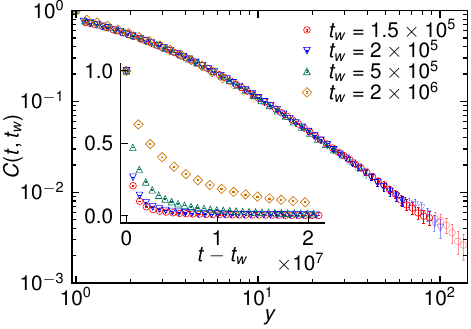}}
\subcaptionbox{$\sigma=0.8$ \label{0.8}}{\includegraphics[width=0.64\columnwidth]{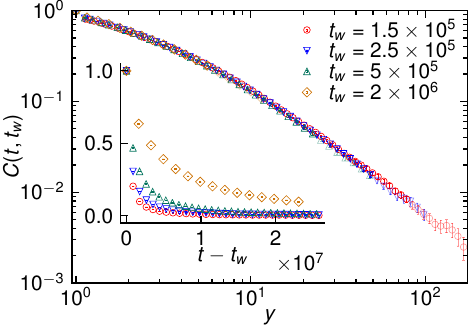}}
\subcaptionbox{$\sigma=0.9$ \label{0.9}}{\includegraphics[width=0.64\columnwidth]{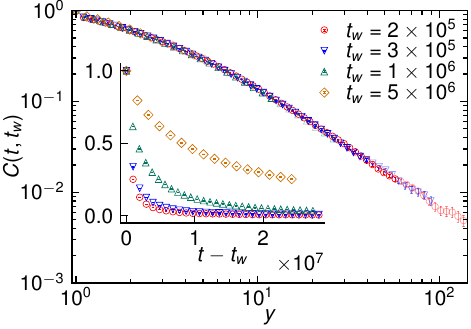}}\\
\subcaptionbox{$\sigma=1$ \label{1}}{\includegraphics[width=0.64\columnwidth]{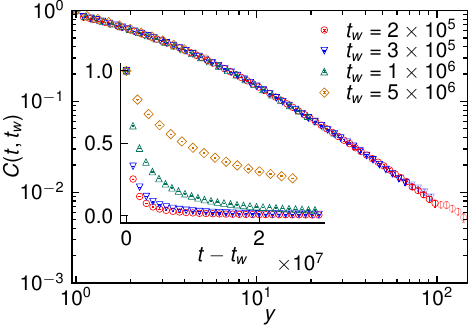}}
\subcaptionbox{$\sigma=1.1$ \label{1.1}}{\includegraphics[width=0.64\columnwidth]{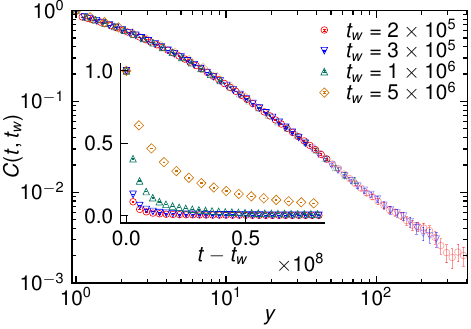}}
\subcaptionbox{$\sigma=1.5$ \label{1.5}}{\includegraphics[width=0.64\columnwidth]{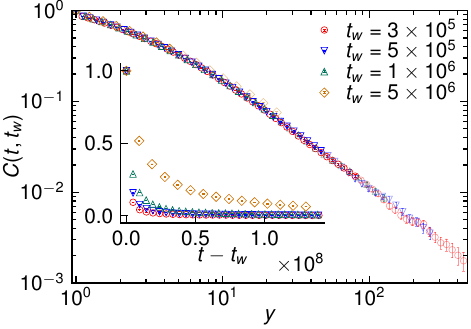}}\\
\subcaptionbox{$\sigma=2$ \label{2}}{\includegraphics[width=0.64\columnwidth]{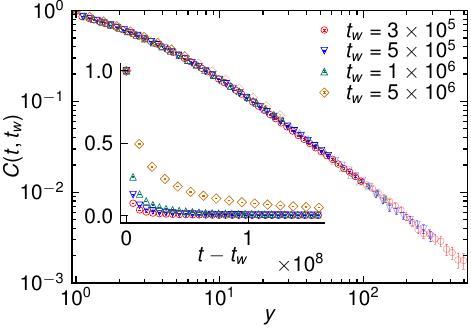}}
\subcaptionbox{$\sigma=3$ \label{3}}{\includegraphics[width=0.64\columnwidth]{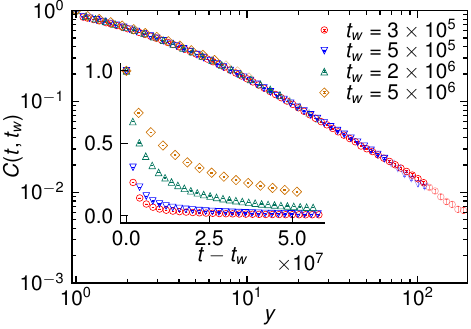}}
\subcaptionbox{$\sigma=10$ \label{10}}{\includegraphics[width=0.64\columnwidth]{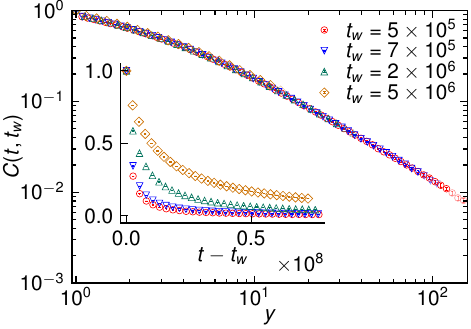}}\\
\subcaptionbox{NN \label{NN}}{\includegraphics[width=0.64\columnwidth]{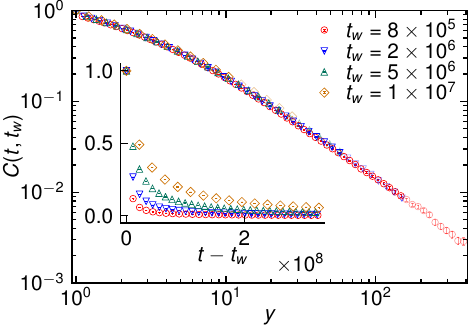}}
\caption{\raggedright Illustration of dynamical scaling with respect to $y=t/t_w$ (main plots) and loss of time-translation invariance (insets).}
\label{fig:time-translation-invariance}
\end{figure*}

One main aspect of aging is the loss of time-translation invariance of the autocorrelation function $C(t,t_w)$, i.e., $C(t,t_w)$ shows no scaling as a function of $t-t_w$.
Instead one observes dynamical scaling in terms of $y=t/t_w$ or $x=\ell(t)/\ell(t_w)$.
For the scaling variable $y$, this is demonstrated in Fig.~\ref{fig:time-translation-invariance}, where for all the considered values of $\sigma$ and the nearest neighbor model (NN) the autocorrelation function $C(t,t_w)$ is shown for different waiting times.
The main plots show $C(t,t_w)$ as a function of $y$ and the insets show the same data as a function of the $t-t_w$.
The plots contain also data which is finite-size affected (see next section) for which deviations from the master curve are expected.
For a facilitated distinction of the finite-size affected data, we have plotted the corresponding points transparently.

While clearly time-translation invariance is broken, as visible from the inset, the data collapse in the main plots is excellent for all $\sigma$ and the NN model.

The earliest waiting time was chosen in such a way that it is minimal but without affecting the quality of the dynamical scaling.
For the fits in the next section, only the data for this earliest $t_w$ is taken since it offers the longest period of finite-size unaffected data.
The other waiting times were chosen in such a way that dynamical scaling is clearly visible.

\begin{figure*}
\subcaptionbox{$\sigma=0.6$ \label{0.6}}{\includegraphics[width=0.64\columnwidth]{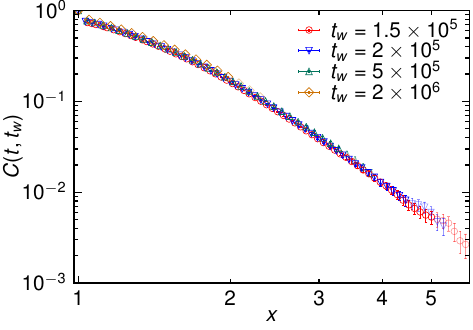}}
\subcaptionbox{$\sigma=0.8$ \label{0.8}}{\includegraphics[width=0.64\columnwidth]{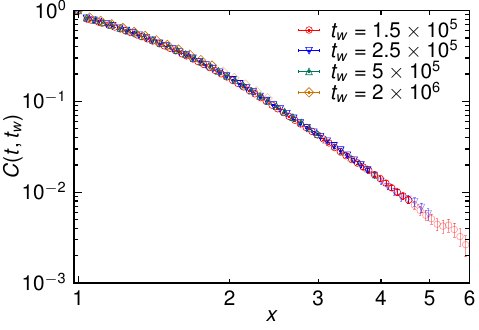}}
\subcaptionbox{$\sigma=0.9$ \label{0.9}}{\includegraphics[width=0.64\columnwidth]{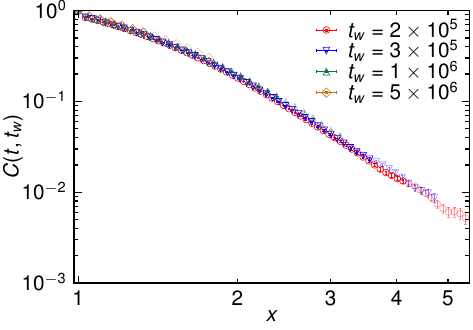}}\\
\subcaptionbox{$\sigma=1$ \label{1}}{\includegraphics[width=0.64\columnwidth]{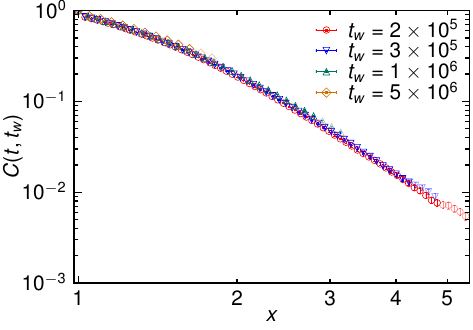}}
\subcaptionbox{$\sigma=1.1$ \label{1.1}}{\includegraphics[width=0.64\columnwidth]{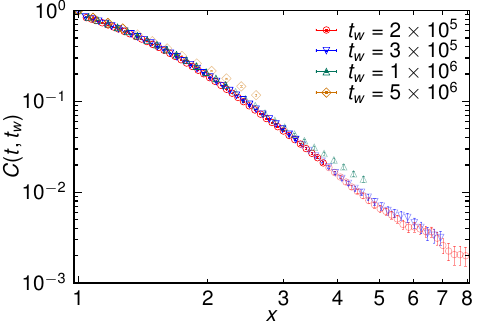}}
\subcaptionbox{$\sigma=1.5$ \label{1.5}}{\includegraphics[width=0.64\columnwidth]{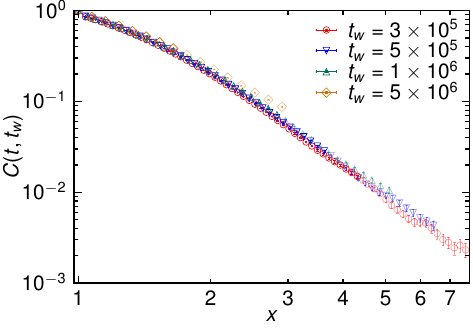}}\\
\subcaptionbox{$\sigma=2$ \label{2}}{\includegraphics[width=0.64\columnwidth]{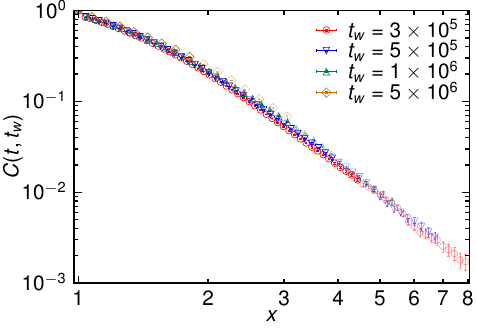}}
\subcaptionbox{$\sigma=3$ \label{3}}{\includegraphics[width=0.64\columnwidth]{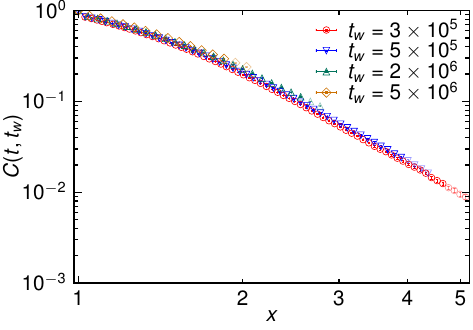}}
\subcaptionbox{$\sigma=10$ \label{10}}{\includegraphics[width=0.64\columnwidth]{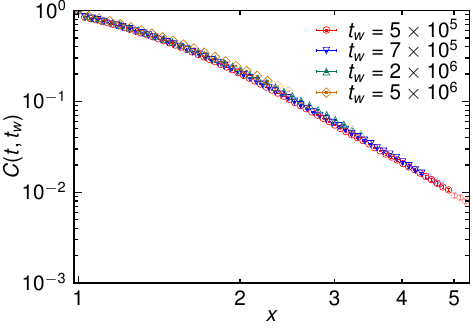}}\\
\subcaptionbox{NN \label{NN}}{\includegraphics[width=0.64\columnwidth]{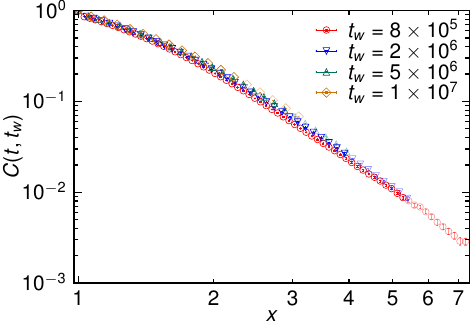}}
\caption{\raggedright Illustration of dynamical scaling with respect to $x=\ell(t)/\ell(t_w)$. For most $\sigma$ the data collapse is considerably worse than for the scaling variable $y$ used in Fig.~\ref{fig:time-translation-invariance}.}
\label{fig:dynamical-scaling-x}
\end{figure*}

Using the same values of $t_w$ as in Fig.~\ref{fig:time-translation-invariance}, the quality of the dynamical scaling for the variable $x$ is investigated in Fig.~\ref{fig:dynamical-scaling-x}, where another illustration of the broken time-translation invariance is not provided but the finite-size affected data are again plotted transparently.
With an appropriate zoom into the figure, it is clearly visible, that the quality of the dynamical scaling for most values of $\sigma$ is noticeably compromised.
Due to the more consistent dynamical scaling in $y$ we will only perform the fits using $y$ as scaling variable.

\section*{Fitting procedure}
Now that the prerequisites for aging are checked, the autocorrelation exponent can be extracted from the data.
In the following, we present a detailed discussion of the fitting procedure for the two different ans\"atze where first a simple (asymptotic) power-law
\begin{equation}
  \label{eq:asymptotic}
  f(y) = Ay^{-\lambda/z}
\end{equation}
is used.
Additionally, an ansatz 
\begin{equation}
  \label{eq:correction}
  f(y) = Ay^{-\lambda/z}(1-B/y)
\end{equation}
is considered which includes a low-order correction term, accounting for pre-asymptotic effects.
The growth exponent $z$ is always kept fixed to its theoretical value $z=\min(2+\sigma, 3)$.
At the crossover point $\sigma=1$ both the equilibrium critical behavior as well as the asymptotic growth of the characteristic length scale $\ell(t)$ carry logarithmic corrections.
Since the autocorrelation function $C(t,t_w)$ for $\sigma=1$ does not show any peculiarities indicating a possible logarithmic correction, however, we do not include any logarithmic correction term into the fitting ansatz for $\sigma=1$ either.
\par
\begin{figure*}
\subcaptionbox{$\sigma=0.6$ \label{0.6}}{\includegraphics[width=0.64\columnwidth]{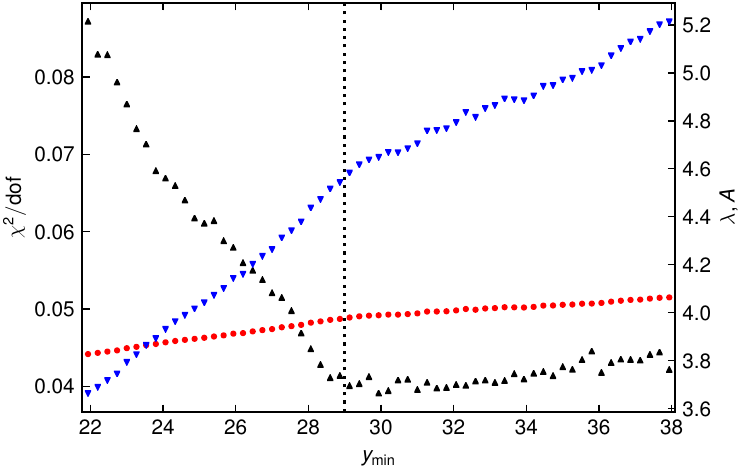}}
\subcaptionbox{$\sigma=0.8$ \label{0.8}}{\includegraphics[width=0.64\columnwidth]{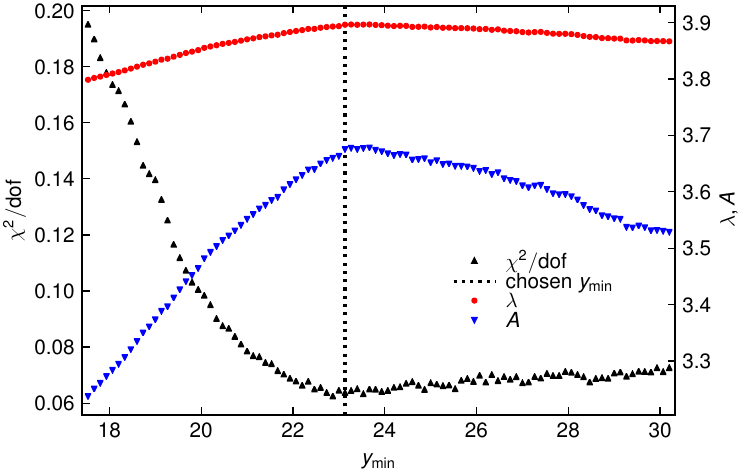}}
\subcaptionbox{$\sigma=0.9$ \label{0.9}}{\includegraphics[width=0.64\columnwidth]{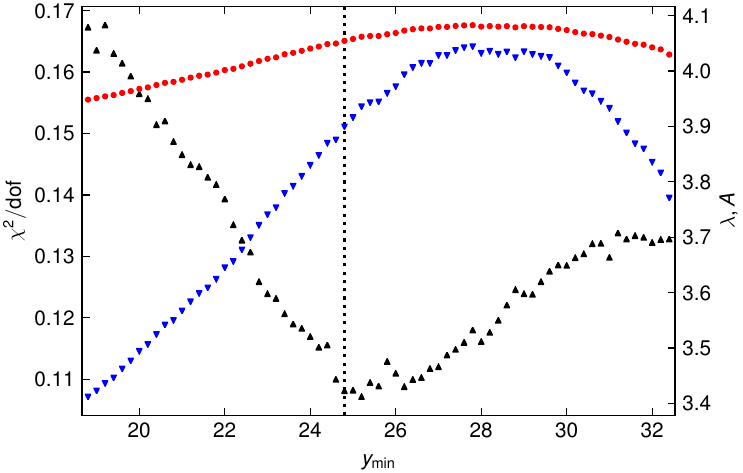}}\\
\subcaptionbox{$\sigma=1$ \label{1}}{\includegraphics[width=0.64\columnwidth]{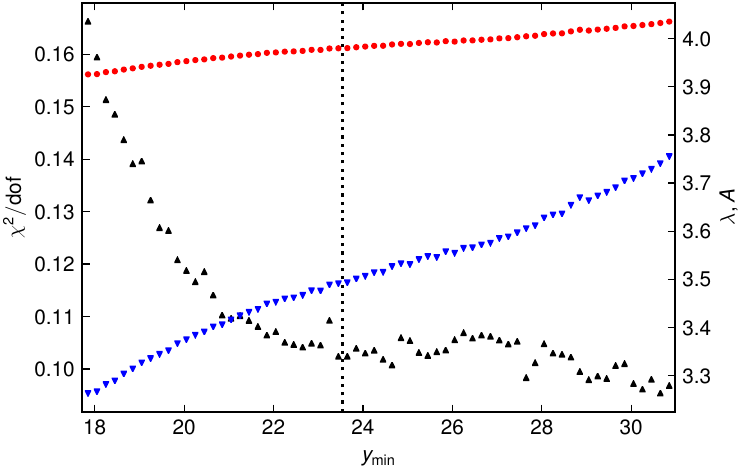}}
\subcaptionbox{$\sigma=1.1$ \label{1.1}}{\includegraphics[width=0.64\columnwidth]{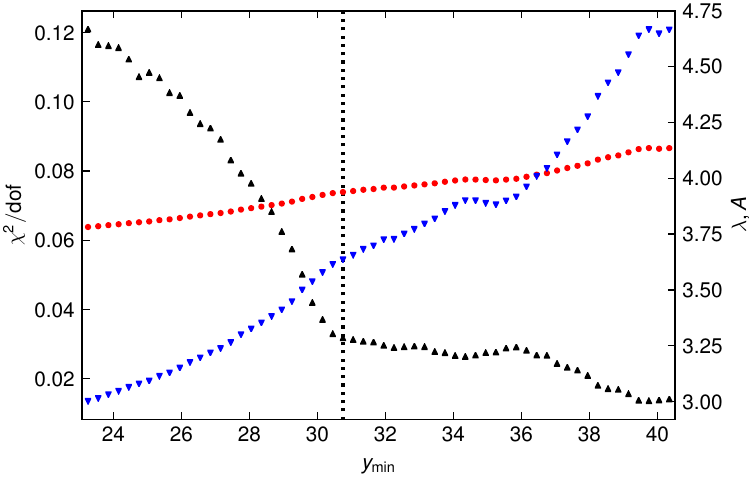}}
\subcaptionbox{$\sigma=1.5$ \label{1.5}}{\includegraphics[width=0.64\columnwidth]{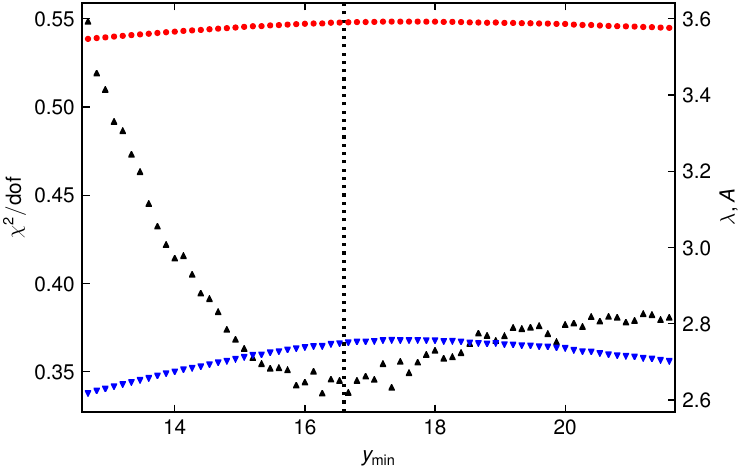}}\\
\subcaptionbox{$\sigma=2$ \label{2}}{\includegraphics[width=0.64\columnwidth]{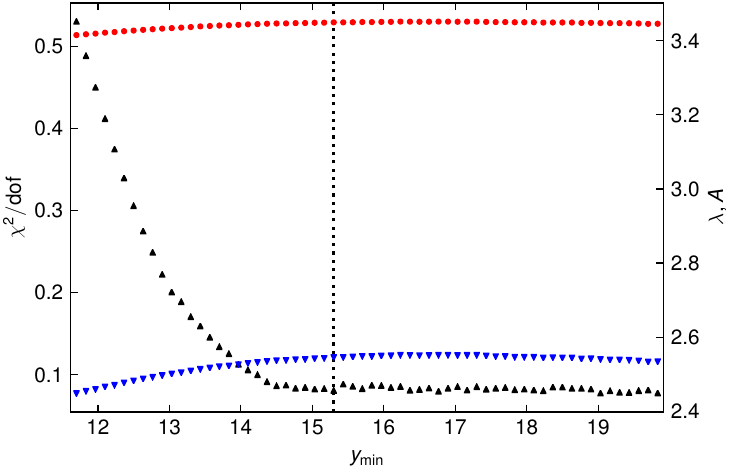}}
\subcaptionbox{$\sigma=3$ \label{3}}{\includegraphics[width=0.64\columnwidth]{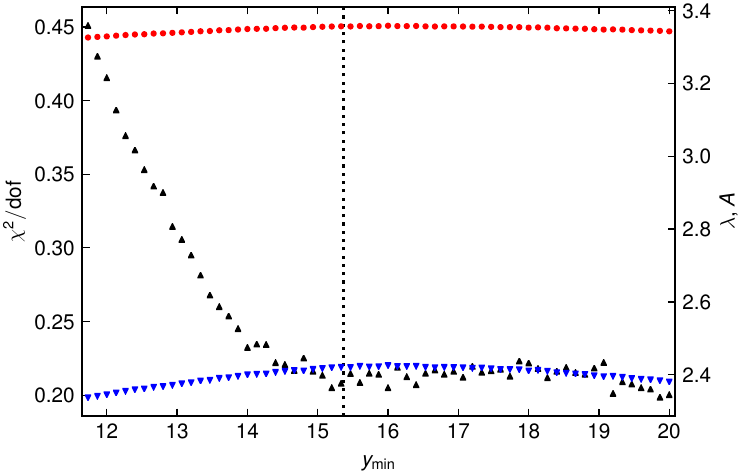}}
\subcaptionbox{$\sigma=10$ \label{10}}{\includegraphics[width=0.64\columnwidth]{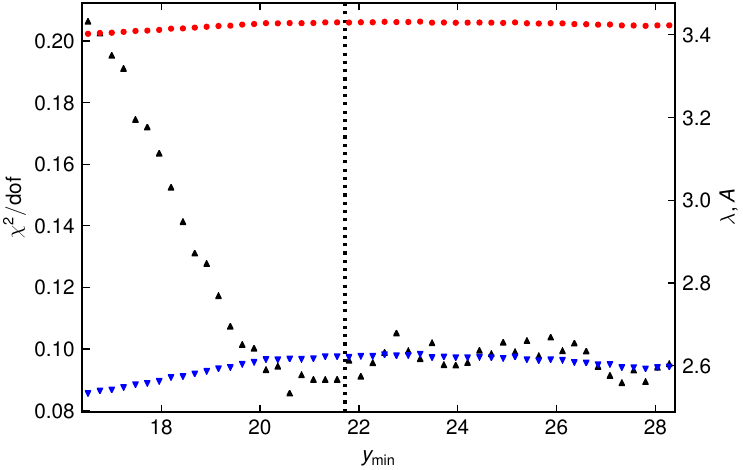}}\\
\subcaptionbox{NN \label{100}}{\includegraphics[width=0.64\columnwidth]{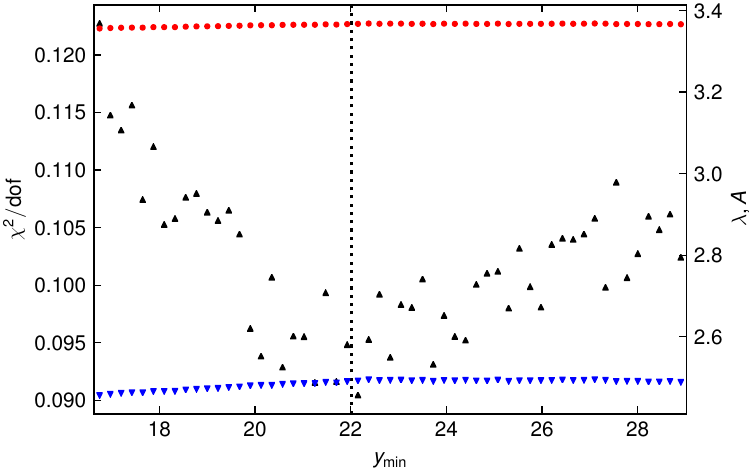}}
\caption{\raggedright Illustration of the influence of the choice of the lower fit boundary $y_\mathrm{min}$ on $\chi^2/\mathrm{dof}$ and the resulting fit parameters for the asymptotic fit ansatz $f(y) = Ay^{-\lambda/z}$ with $y=t/t_w$ using the smallest $t_w$ for each $\sigma$ (cf. Fig.~\ref{fig:time-translation-invariance}).
  The vertical dotted lines indicate the finally chosen $y_\mathrm{min}$.}
\label{fig:ymin-asymptotic}
\end{figure*}
\begin{figure*}
\subcaptionbox{$\sigma=0.6$ \label{0.6}}{\includegraphics[width=0.64\columnwidth]{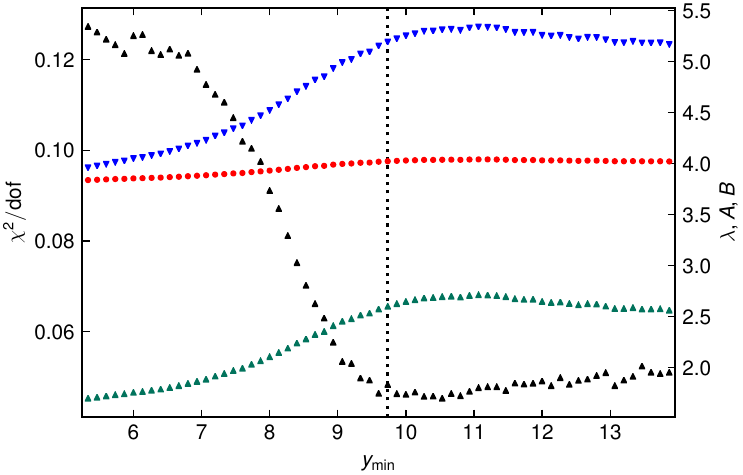}}
\subcaptionbox{$\sigma=0.8$ \label{0.8}}{\includegraphics[width=0.64\columnwidth]{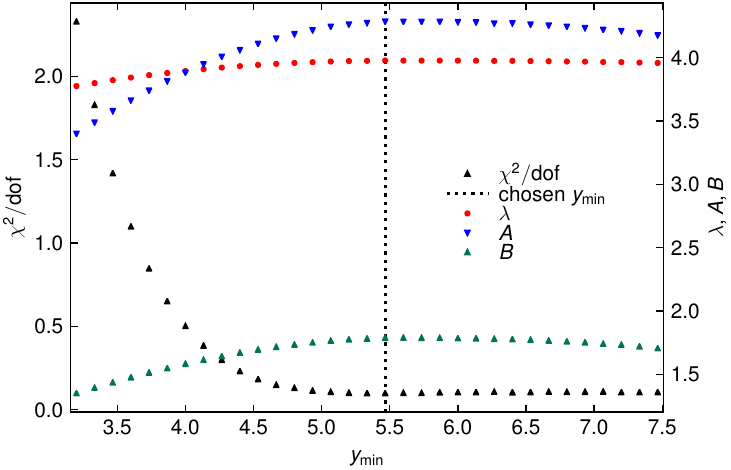}}
\subcaptionbox{$\sigma=0.9$ \label{0.9}}{\includegraphics[width=0.64\columnwidth]{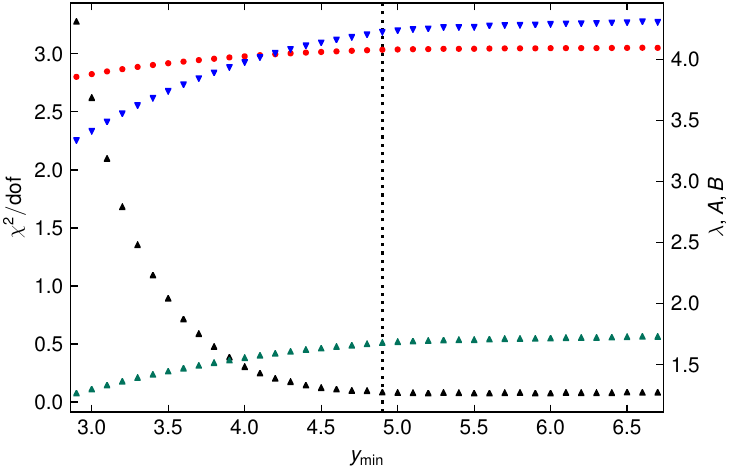}}\\
\subcaptionbox{$\sigma=1$ \label{1}}{\includegraphics[width=0.64\columnwidth]{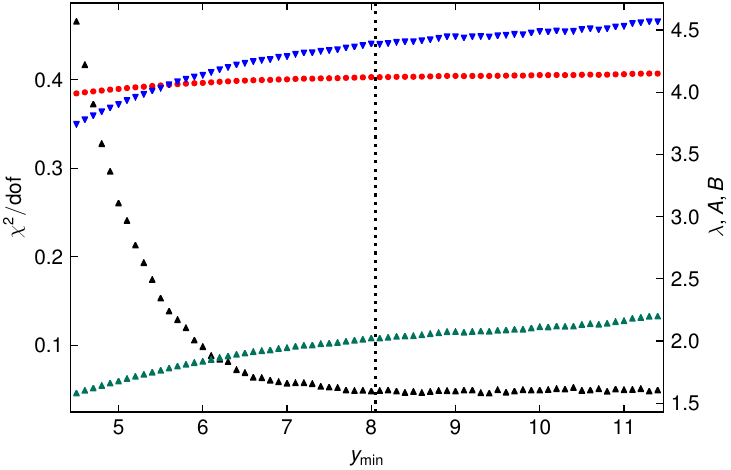}}
\subcaptionbox{$\sigma=1.1$ \label{1.1}}{\includegraphics[width=0.64\columnwidth]{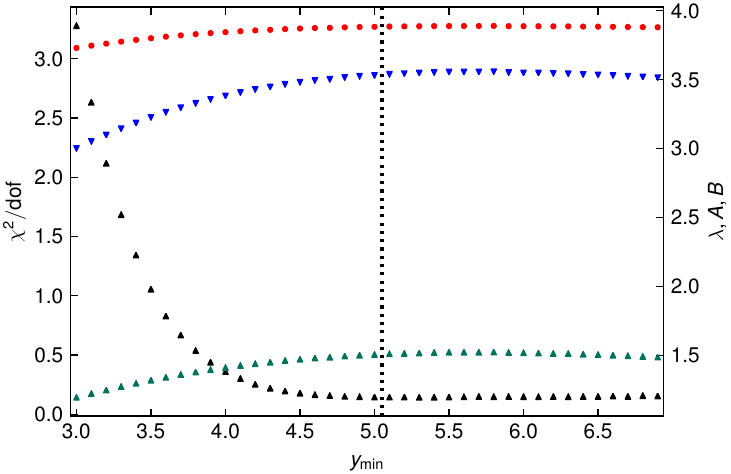}}
\subcaptionbox{$\sigma=1.5$ \label{1.5}}{\includegraphics[width=0.64\columnwidth]{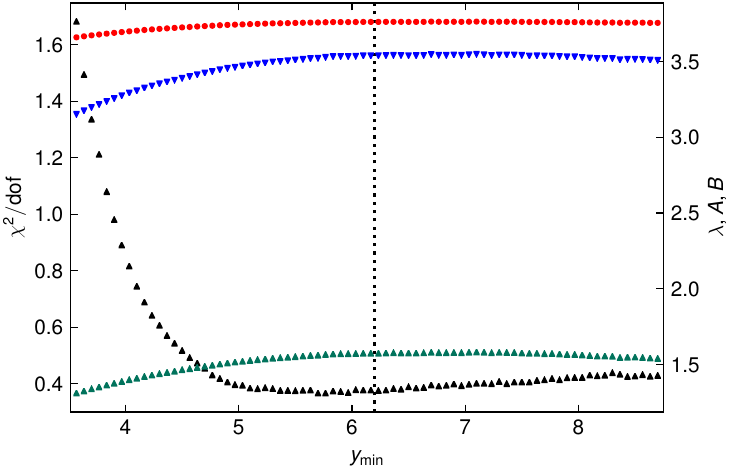}}\\
\subcaptionbox{$\sigma=2$ \label{2}}{\includegraphics[width=0.64\columnwidth]{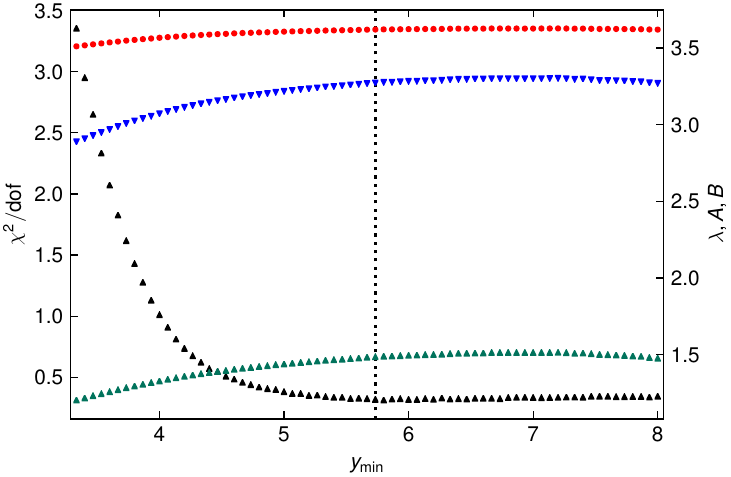}}
\subcaptionbox{$\sigma=3$ \label{3}}{\includegraphics[width=0.64\columnwidth]{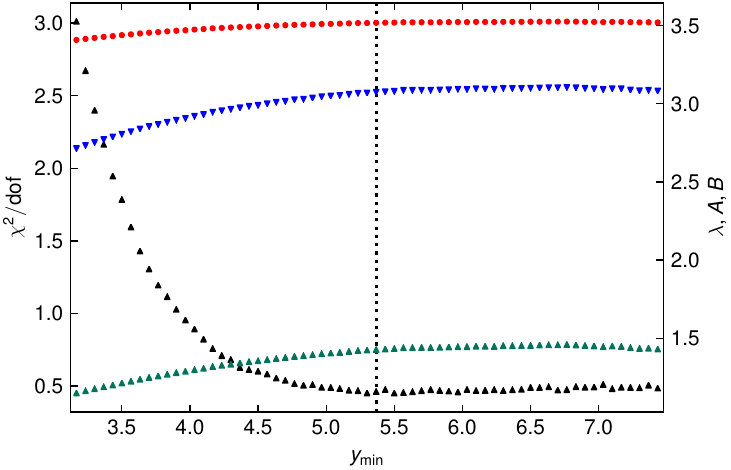}}
\subcaptionbox{$\sigma=10$ \label{10}}{\includegraphics[width=0.64\columnwidth]{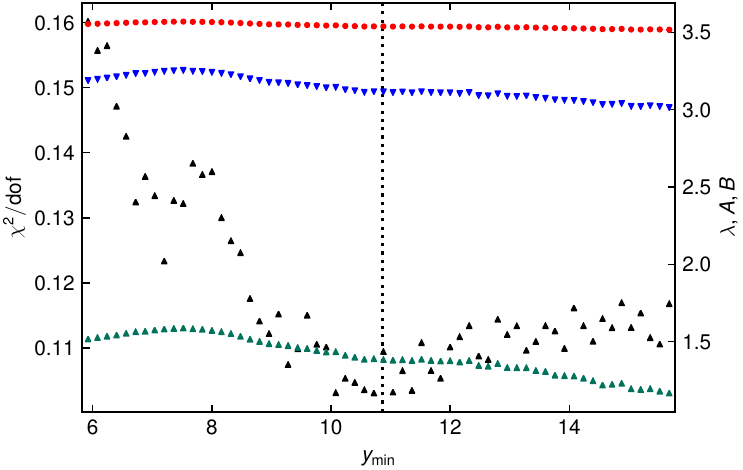}}\\
\subcaptionbox{NN \label{100}}{\includegraphics[width=0.64\columnwidth]{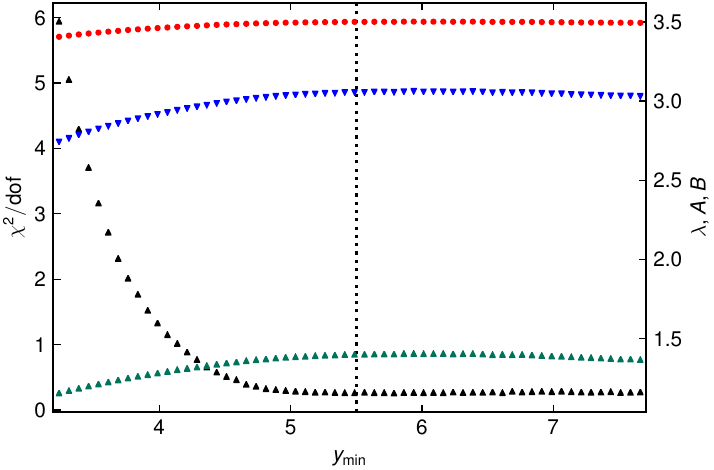}}
\caption{\raggedright Same as Fig.~\ref{fig:ymin-asymptotic} for the fit ansatz $f(y) = Ay^{-\lambda/z}(1-B/y)$ with correction term.}
\label{fig:ymin-correction}
\end{figure*}

As discussed in the main text, the fitting interval $[y_\mathrm{min},y_\mathrm{max}]$ is chosen in the following way:
First all data for which noticeable finite-size effects in $\ell(t)$ are visible (cf.\ Fig.~2 in the main text) are discarded.
The $y_\mathrm{max}$ is determined roughly from the deviations of the autocorrelation functions for the chosen $t_w$ and the next larger $t_w$.
In some cases, however, the fluctuation of $C(t,t_w)$ are already quite strong, even for $y<y_\mathrm{max}$, in which cases we reduced the value of $y_\mathrm{max}$ further.
This avoids an increase in statistical errors and the accumulation of systematic errors.
The values of $y_\mathrm{min}$ were chosen in such a way that the fit interval is maximal with the restriction that no strong systematic trend in the resulting fit parameters is visible and the value of $\chi^2/\mathrm{dof}$ does not show any strong increase.
Note that the value of $\chi^2/\mathrm{dof}$ has no absolute meaning since the data are correlated.
However, a strong ramp remains an indication that the ansatz becomes inappropriate for the data.
The above procedure is illustrated in Figs.~\ref{fig:ymin-asymptotic} and~\ref{fig:ymin-correction} for the ansatz without and with correction term, respectively, showing the dependence of $\chi^2/\mathrm{dof}$ and the fitting parameters on $y_\mathrm{min}$.
Both ans\"atze show a sharp increase of $\chi^2/\mathrm{dof}$ for decreasing $y_\mathrm{min}$, which is accompanied also by an onset of stronger trends in the resulting fit parameters.
Especially for the fits including the correction term in Fig.~\ref{fig:ymin-correction}, there are very pronounced plateaus in the resulting fit parameters which underscore the robustness of the ansatz.
We have thus decided to present the values of $\lambda$ obtained from the fits with the correction term in the main text, since they seem to be more robust and thus less prone to systematic errors.
\par
\begin{figure*}
\subcaptionbox{$\sigma=0.6,\ \lambda=4.03\pm0.38$ \label{0.6}}{\includegraphics[width=0.64\columnwidth]{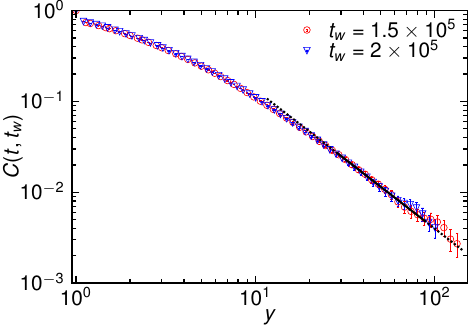}}
\subcaptionbox{$\sigma=0.8,\ \lambda=3.93\pm0.16$ \label{0.8}}{\includegraphics[width=0.64\columnwidth]{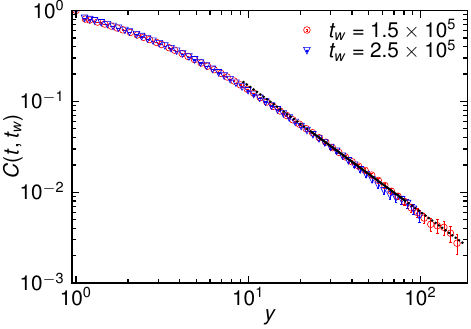}}
\subcaptionbox{$\sigma=0.9,\ \lambda=3.86\pm0.09$ \label{0.9}}{\includegraphics[width=0.64\columnwidth]{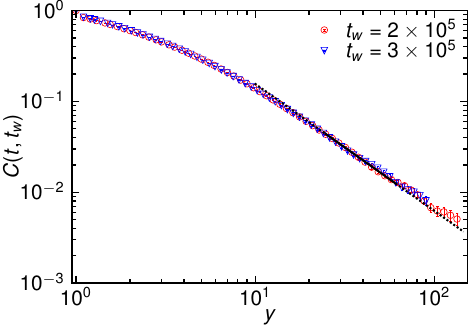}}\\
\subcaptionbox{$\sigma=1,\ \lambda=3.97\pm0.13$ \label{1}}{\includegraphics[width=0.64\columnwidth]{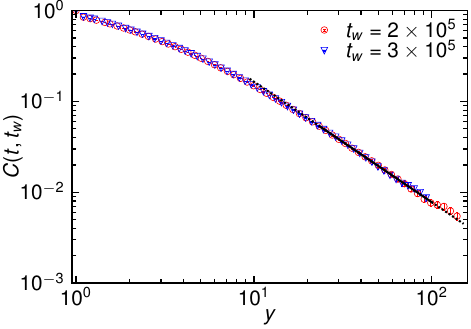}}
\subcaptionbox{$\sigma=1.1,\ \lambda=4.02\pm0.15$ \label{1.1}}{\includegraphics[width=0.64\columnwidth]{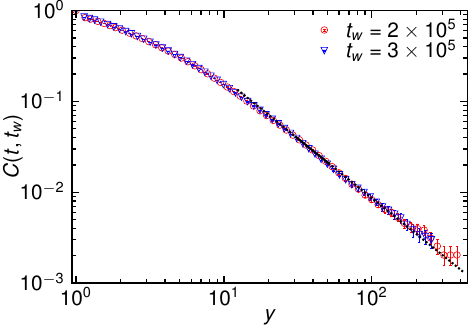}}
\subcaptionbox{$\sigma=1.5,\ \lambda=3.58\pm0.06$ \label{1.5}}{\includegraphics[width=0.64\columnwidth]{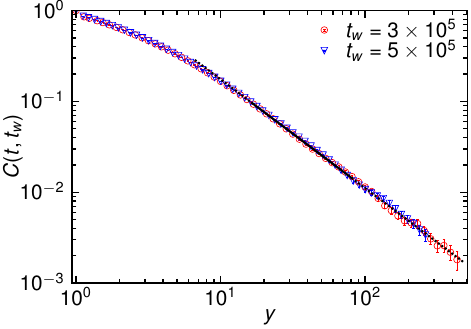}}\\
\subcaptionbox{$\sigma=2,\ \lambda=3.47\pm0.04$ \label{2}}{\includegraphics[width=0.64\columnwidth]{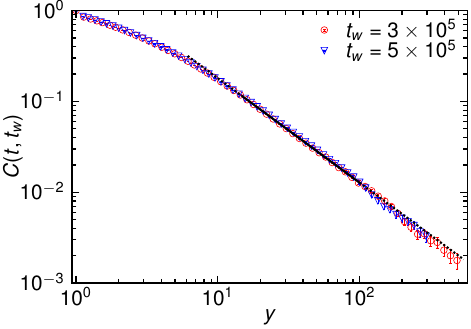}}
\subcaptionbox{$\sigma=3,\ \lambda=3.36\pm0.04$ \label{3}}{\includegraphics[width=0.64\columnwidth]{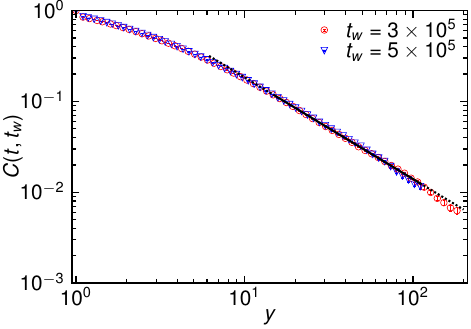}}
\subcaptionbox{$\sigma=10,\ \lambda=3.43\pm0.06$ \label{10}}{\includegraphics[width=0.64\columnwidth]{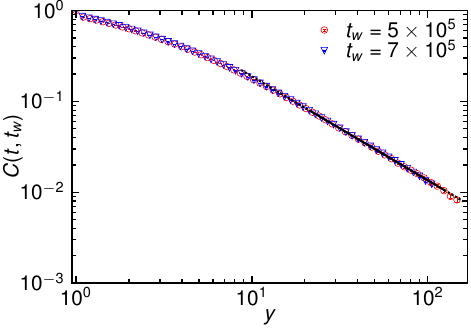}}\\
\subcaptionbox{NN,\ $\lambda=3.45\pm0.08$ \label{NN}}{\includegraphics[width=0.64\columnwidth]{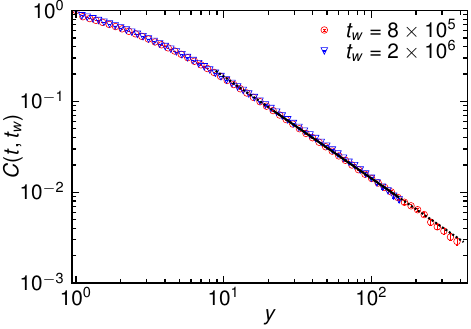}}
\caption{\raggedright Plots of the asymptotic fits $f(y) = Ay^{-\lambda/z}$ to $C(t, t_w)$ where the data for the smaller $t_w$ are fitted in the previously chosen intervals $[y_\mathrm{min}, y_\mathrm{max}]$.
  The solid lines indicate the fitting range and the dotted lines represent the continuation of the fits beyond the fitting interval.}
\label{fig:fits-asymptotic}
\end{figure*}
\begin{figure*}
\subcaptionbox{$\sigma=0.6,\ \lambda=4.06\pm0.26$ \label{0.6}}{\includegraphics[width=0.64\columnwidth]{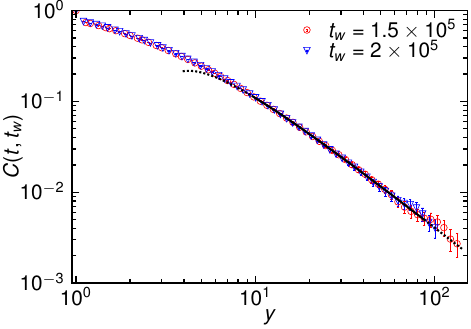}}
\subcaptionbox{$\sigma=0.8,\ \lambda=3.98\pm0.09$ \label{0.8}}{\includegraphics[width=0.64\columnwidth]{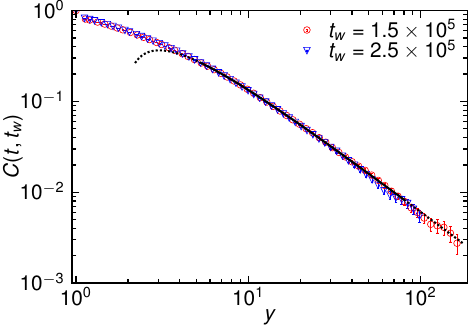}}
\subcaptionbox{$\sigma=0.9,\ \lambda=4.06\pm0.07$ \label{0.9}}{\includegraphics[width=0.64\columnwidth]{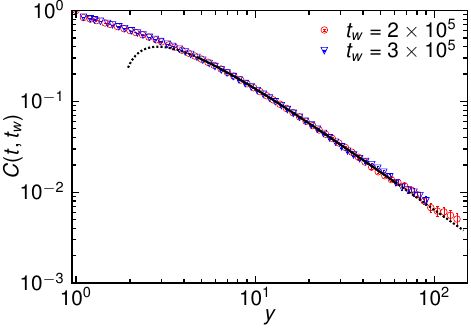}}\\
\subcaptionbox{$\sigma=1,\ \lambda=4.12\pm0.10$ \label{1}}{\includegraphics[width=0.64\columnwidth]{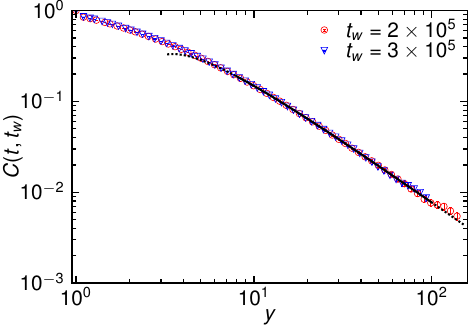}}
\subcaptionbox{$\sigma=1.1,\ \lambda=3.93\pm0.06$ \label{1.1}}{\includegraphics[width=0.64\columnwidth]{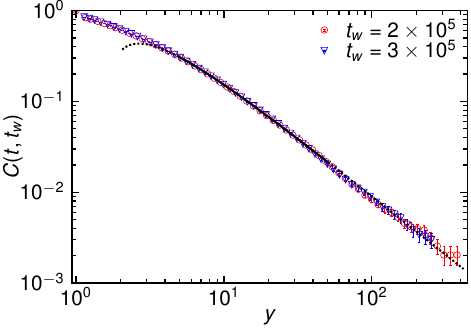}}
\subcaptionbox{$\sigma=1.5,\ \lambda=3.74\pm0.05$ \label{1.5}}{\includegraphics[width=0.64\columnwidth]{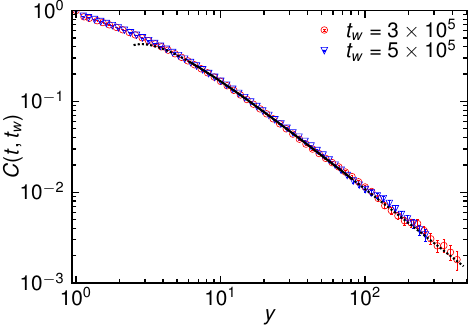}}\\
\subcaptionbox{$\sigma=2,\ \lambda=3.62\pm0.04$ \label{2}}{\includegraphics[width=0.64\columnwidth]{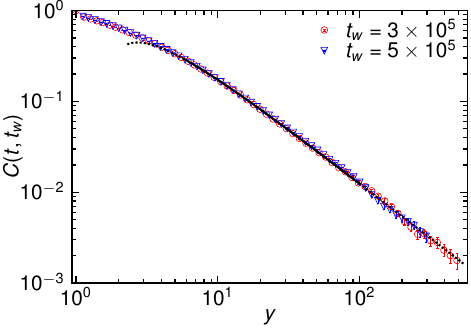}}
\subcaptionbox{$\sigma=3,\ \lambda=3.52\pm0.04$ \label{3}}{\includegraphics[width=0.64\columnwidth]{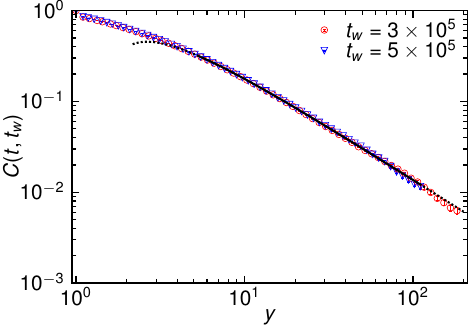}}
\subcaptionbox{$\sigma=10,\ \lambda=3.54\pm0.08$ \label{10}}{\includegraphics[width=0.64\columnwidth]{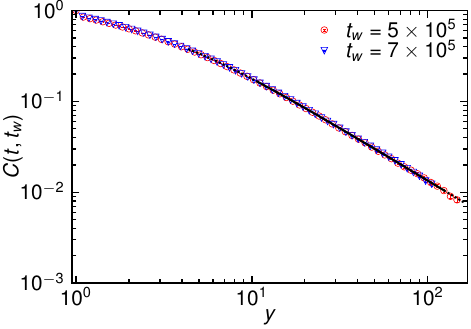}}\\
\subcaptionbox{NN,\ $\lambda=3.50\pm0.03$ \label{100}}{\includegraphics[width=0.64\columnwidth]{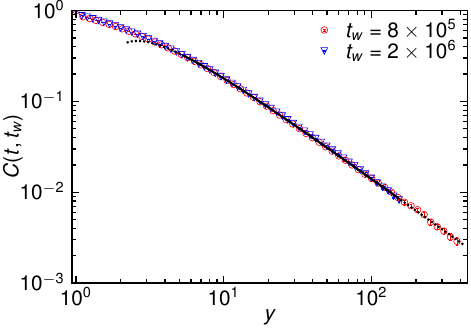}}
\caption{\raggedright Same as Fig.~\ref{fig:fits-asymptotic} for the fits $f(y) = Ay^{-\lambda/z}(1-B/y)$ with correction term.}
\label{fig:fits-correction}
\end{figure*}
\begin{figure*}
  \includegraphics{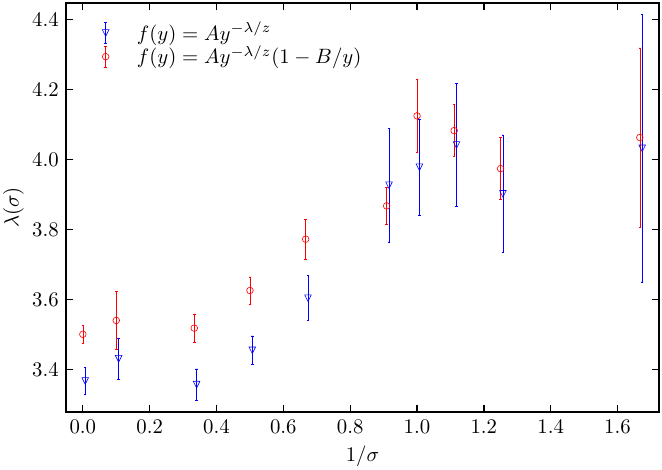}
  \caption{\raggedright Comparison of the $\sigma$-dependence of $\lambda$ for the two considered fit ans\"atze.}
  \label{fig:direct-comparison}
\end{figure*}

The resulting fits, together with the fitted correlation function are presented in Fig.~\ref{fig:fits-asymptotic} and Fig.~\ref{fig:fits-correction} without and with correction term, respectively.
The fits with correction term describe the data over a significantly longer interval compared to the corresponding fit without correction, explaining the enhanced stability of the resulting fit parameters.
In Tables~\ref{tab:asymptotic} and~\ref{tab:correction} the input parameters for the fits and the parameters of the outcoming best fitting curves are reported for convenience such that the interested reader can better compare the different settings and results.
Finally, the dependence of $\lambda$ on $\sigma$ is compared for the two ans\"atze in Fig.~\ref{fig:direct-comparison}.
There it is clearly visible that the fits with and without correction are for most values of $\sigma$ in very good qualitative agreement.
The statistical errors which we assess by Jackknifing over the different realizations (initial conditions and thermal noises) tend to be reduced upon introducing the correction term.
While increased statistical errors would be expected due to the additional fitting parameter, the significantly extended fitting ranges overcompensate this effect, enabling a clear statement about the crossover behavior in the regime $1 \lesssim  \sigma  \lesssim 2$.

\begin{table*}
  \caption{\raggedright Fit parameters for the asymptotic fit $f(y) = Ay^{-\lambda/z}$.}
            
\label{tab:asymptotic}
\begin{tabular}{llllllll|lll}
  \hline\hline
  \multicolumn{8}{c|}{input} & \multicolumn{3}{c}{results}\\\hline
  \multicolumn{1}{c}{$L$} & \multicolumn{1}{c}{$\sigma$} & \multicolumn{1}{c}{$t_w$} & \multicolumn{1}{c}{$t_\mathrm{min}$} & \multicolumn{1}{c}{$t_\mathrm{max}$} & \multicolumn{1}{c}{$\ell(t_w)$} & \multicolumn{1}{c}{$\ell(t_\mathrm{min})$} & \multicolumn{1}{c|}{$\ell(t_\mathrm{max})$} & \multicolumn{1}{c}{$\chi^2/\mathrm{dof}$} & \multicolumn{1}{c}{$\lambda$} & \multicolumn{1}{c}{$A$}\\\hline
  2048 & \hphantom{0}0.6 &  $1.50 \times 10^{5}$  &  $4.35 \times 10^{6}$  &  $1.32 \times 10^{7}$  & 60.2 & 200.1 & 304 & 0.04 &  $4.03(39)$  &  $4.9(2.6)$   \\
2048 & \hphantom{0}0.8 &  $1.50 \times 10^{5}$  &  $3.47 \times 10^{6}$  &  $1.22 \times 10^{7}$  & 46.8 & 136.9 & 214 & 0.06 &  $3.90(17)$  &  $3.70(74)$   \\
2048 & \hphantom{0}0.9 &  $2.00 \times 10^{5}$  &  $4.96 \times 10^{6}$  &  $1.22 \times 10^{7}$  & 46.2 & 137.9 & 191 & 0.11 &  $4.04(18)$  &  $3.83(79)$   \\
2048 & \hphantom{0}1 &  $2.00 \times 10^{5}$  &  $4.71 \times 10^{6}$  &  $2.00 \times 10^{7}$  & 42.2 & 122.2 & 204 & 0.11 &  $3.98(14)$  &  $3.49(55)$   \\
1024 & \hphantom{0}1.1 &  $2.00 \times 10^{5}$  &  $6.15 \times 10^{6}$  &  $1.02 \times 10^{7}$  & 38.9 & 121.8 & 145 & 0.03 &  $3.93(17)$  &  $3.60(69)$   \\
1024 & \hphantom{0}1.5 &  $3.00 \times 10^{5}$  &  $4.98 \times 10^{6}$  &  $2.43 \times 10^{7}$  & 34.4 & \hphantom{0}86.3 & 148 & 0.34 &  $3.604(64)$  &  $2.79(19)$   \\
1024 & \hphantom{0}2 &  $3.00 \times 10^{5}$  &  $4.59 \times 10^{6}$  &  $3.03 \times 10^{7}$  & 28.3 & \hphantom{0}68.4 & 129 & 0.08 &  $3.455(42)$  &  $2.56(11)$   \\
\hphantom{0}512 & \hphantom{0}3 &  $3.00 \times 10^{5}$  &  $4.61 \times 10^{6}$  &  $3.33 \times 10^{7}$  & 23.3 & \hphantom{0}55.7 & 109 & 0.21 &  $3.357(45)$  &  $2.42(11)$   \\
\hphantom{0}512 & 10 &  $5.00 \times 10^{5}$  &  $1.09 \times 10^{7}$  &  $6.25 \times 10^{7}$  & 20.6 & \hphantom{0}55.7 & 101 & 0.09 &  $3.431(59)$  &  $2.62(18)$   \\
1024 & NN &  $8.00 \times 10^{5}$  &  $1.76 \times 10^{7}$  &  $1.21 \times 10^{8}$  & 20.4 & \hphantom{0}56.0 & 108 & 0.09 &  $3.367(39)$  &  $2.49(11)$   \\\hline\hline
\end{tabular}
\end{table*}

\begin{table*}
  \caption{\raggedright Fit parameters for the fit with correction $f(y) = Ay^{-\lambda/z}(1-B/y)$.}
            
\label{tab:correction}

\begin{tabular}{llllllll|llll}
  
  \hline\hline
  
  \multicolumn{8}{c|}{input} & \multicolumn{4}{c}{results}\\\hline
  \multicolumn{1}{c}{$L$} & \multicolumn{1}{c}{$\sigma$} & \multicolumn{1}{c}{$t_w$} & \multicolumn{1}{c}{$t_\mathrm{min}$} & \multicolumn{1}{c}{$t_\mathrm{max}$} & \multicolumn{1}{c}{$\ell(t_w)$} & \multicolumn{1}{c}{$\ell(t_\mathrm{min})$} & \multicolumn{1}{c|}{$\ell(t_\mathrm{max})$} & \multicolumn{1}{c}{$\chi^2/\mathrm{dof}$} & \multicolumn{1}{c}{$\lambda$} & \multicolumn{1}{c}{$A$} & \multicolumn{1}{c}{$B$}\\\hline
  2048 & \hphantom{0}0.6 &  $1.50 \times 10^{5}$  &  $1.46 \times 10^{6}$  &  $1.32 \times 10^{7}$  & 60.2 & 134.1 & 304 & 0.05 &  $4.06(26)$  &  $5.5(1.9)$  &  $2.71(88)$\\
2048 & \hphantom{0}0.8 &  $1.50 \times 10^{5}$  &  $8.20 \times 10^{5}$  &  $1.22 \times 10^{7}$  & 46.8 & \hphantom{0}82.7 & 214 & 0.10 &  $3.974(89)$  &  $4.26(41)$  &  $1.77(16)$\\
2048 & \hphantom{0}0.9 &  $2.00 \times 10^{5}$  &  $9.80 \times 10^{5}$  &  $1.22 \times 10^{7}$  & 46.2 & \hphantom{0}78.2 & 191 & 0.09 &  $4.082(75)$  &  $4.24(30)$  &  $1.68(11)$\\
2048 & \hphantom{0}1 &  $2.00 \times 10^{5}$  &  $1.61 \times 10^{6}$  &  $2.00 \times 10^{7}$  & 42.2 & \hphantom{0}84.3 & 204 & 0.05 &  $4.12(11)$  &  $4.42(53)$  &  $2.06(31)$\\
1024 & \hphantom{0}1.1 &  $2.00 \times 10^{5}$  &  $1.01 \times 10^{6}$  &  $1.02 \times 10^{7}$  & 38.9 & \hphantom{0}65.9 & 145 & 0.14 &  $3.867(53)$  &  $3.49(18)$  &  $1.487(89)$\\
1024 & \hphantom{0}1.5 &  $3.00 \times 10^{5}$  &  $1.86 \times 10^{6}$  &  $2.43 \times 10^{7}$  & 34.4 & \hphantom{0}62.0 & 148 & 0.38 &  $3.772(58)$  &  $3.58(23)$  &  $1.60(14)$\\
1024 & \hphantom{0}2 &  $3.00 \times 10^{5}$  &  $1.72 \times 10^{6}$  &  $3.03 \times 10^{7}$  & 28.3 & \hphantom{0}49.4 & 129 & 0.32 &  $3.626(40)$  &  $3.29(14)$  &  $1.494(85)$\\
\hphantom{0}512 & \hphantom{0}3 &  $3.00 \times 10^{5}$  &  $1.61 \times 10^{6}$  &  $3.33 \times 10^{7}$  & 23.3 & \hphantom{0}39.4 & 109 & 0.45 &  $3.518(41)$  &  $3.07(13)$  &  $1.424(83)$\\
\hphantom{0}512 & 10 &  $5.00 \times 10^{5}$  &  $5.43 \times 10^{6}$  &  $6.25 \times 10^{7}$  & 20.6 & \hphantom{0}44.2 & 101 & 0.10 &  $3.540(83)$  &  $3.12(34)$  &  $1.39(42)$\\
1024 & NN &  $8.00 \times 10^{5}$  &  $4.40 \times 10^{6}$  &  $1.21 \times 10^{8}$  & 20.4 & \hphantom{0}35.2 & 108 & 0.26 &  $3.500(26)$  &  $3.056(81)$  &  $1.399(54)$\\\hline\hline
\end{tabular}
\end{table*}

